\documentclass[conference,10pt]{IEEEtran}

\usepackage{framed}
\usepackage{tcolorbox}
\usepackage{amsmath}
\usepackage{mathtools}
\usepackage{bbm}
\usepackage[ruled,linesnumbered]{algorithm2e}
\usepackage{bm}
\usepackage{stackrel}
\usepackage{subfig}
\usepackage{epsf}
\usepackage{amsmath,amssymb}

\usepackage{graphicx}
\usepackage{color}
\usepackage{cite}
\usepackage{multirow,tabularx}
\usepackage{hyperref}
\usepackage{ifthen}
\usepackage{epstopdf}
\input{epstopdf.sty}

\usepackage{times}
\input{epsf.sty}

\makeatletter
\def\BState{\State\hskip-\ALG@thistlm}
\makeatother

\newcommand{\hide}[1]{\ifthenelse{\boolean{false}}{#1}{}}

\newtheorem{theorem}{{\bf Theorem}}

\newtheorem{lemma}{{\bf Lemma}}

\newenvironment{definition}[1][Definition]{\begin{trivlist}
\item[\hskip \labelsep {\bfseries #1}]}{\end{trivlist}}

\newcommand{\qed}{\nobreak \ifvmode \relax \else
      \ifdim\lastskip<1.5em \hskip-\lastskip
      \hskip1.5em plus0em minus0.5em \fi \nobreak
      \vrule height0.75em width0.5em depth0.25em\fi}

\newcommand{\beq}{\begin{equation}}
\newcommand{\eeq}{\end{equation}}
\newcommand{\barr}{\begin{array}}
\newcommand{\earr}{\end{array}}

\newcommand{\benum}{\begin{enumerate}}
\newcommand{\eenum}{\end{enumerate}}

\newcommand{\bit}{\begin{itemize}}
\newcommand{\eit}{\end{itemize}}

\newcommand{\bc}{\begin{center}}
\newcommand{\ec}{\end{center}}

\newcommand{\bdes}{\begin{description}}
\newcommand{\edes}{\end{description}}

\newcommand{\bfig}{\begin{figure}}
\newcommand{\efig}{\end{figure}}

\newcommand{\bemq}{\begin{quote} \begin{em}}
\newcommand{\eemq}{\end{em} \end{quote}}

\newcommand{\bmp}{\begin{minipage}}
\newcommand{\emp}{\end{minipage}}

\newcommand{\bsp}{\begin{slide*}}
\newcommand{\esp}{\end{slide*}}
\newcommand{\bsl}{\begin{slide}}
\newcommand{\esl}{\end{slide}}

\newcommand{\blem}{\begin{lemma}}
\newcommand{\elem}{\end{lemma}}
\newcommand{\bthm}{\begin{theorem}}
\newcommand{\ethm}{\end{theorem}}

\IEEEoverridecommandlockouts

\begin{document}

\title{Age Debt: A General Framework For Minimizing Age of Information}
\author{Vishrant Tripathi and Eytan Modiano\\
Laboratory for Information \& Decision Systems, MIT
\thanks{
This work was supported by NSF Grants AST-1547331, CNS-1713725, and CNS-1701964, and by Army Research Office (ARO) grant number W911NF-17-1-0508.
}
}

\IEEEaftertitletext{\vspace{-0.6\baselineskip}}

\maketitle
\begin{abstract}
We consider the problem of minimizing age of information in general single-hop and multihop wireless networks. First, we formulate a way to convert AoI optimization problems into equivalent network stability problems. Then, we propose a heuristic low complexity approach for achieving stability that can handle general network topologies; unicast, multicast and broadcast flows; interference constraints; link reliabilities; and AoI cost functions. We provide numerical results to show that our proposed algorithms behave as well as the best known scheduling and routing schemes available in the literature for a wide variety of network settings.
\end{abstract}

\section{Introduction}
\label{sec:intro}

Many emerging applications require timely delivery of information updates over communication networks. 
Age of Information (AoI) is a metric that captures precisely this notion of timeliness of received information at a destination \cite{kaul2012real,kam2013age,yin17_tit_update_or_wait}. Unlike packet delay, AoI measures the lag in obtaining information at a destination node, and is therefore suited for applications involving time sensitive updates. Age of information, at a destination, is defined as the time that has elapsed since the last received information update was generated at the source. AoI, upon reception of a new update, drops to the time elapsed since generation of the update, and grows linearly otherwise. Over the past few years, there has been a rapidly growing body of work on using AoI as a metric for scheduling policies in wireless networks \cite{kadota2018scheduling, kadota2018scheduling2, talak2018optimizing, maatouk2020optimality, farazi2018age, tripathi2019whittle}. For detailed surveys of AoI literature see \cite{kosta2017age} and \cite{sun2019age_book}.




Scheduling to minimize AoI in single-hop wireless networks has been considered in \cite{kadota2018scheduling,kadota2018scheduling2,talak2018optimizing,maatouk2020optimality}. These works prove constant factor optimality of three classes of policies - randomized, max-weight and Whittle index based; under both reliable and unreliable channels. Further, minimizing general cost functions of AoI in single-hop wireless networks has been considered in \cite{jhun2018age,tripathi2019whittle}. 

AoI minimization over multi-hop wireless networks has also been considered in different settings. In \cite{talak2017minimizing}, the authors developed stationary randomized policies to minimize weighted-sum-AoI of unicast flows with fixed paths in a multi-hop network. AoI minimization in multihop wireless networks with all-to-all broadcast flows was considered in \cite{farazi2018age, farazi2019fundamental}. 

We observe that finding low complexity near optimal scheduling and routing schemes for AoI minimization which handle general network topologies, interference constraints, cost functions, different types of flows and link reliabilities has remained an open problem.

In this work, we develop a unifying framework for making routing and scheduling decisions that optimize AoI in general multihop networks. We do this by transforming AoI-based network optimization problems into network stability problems. Instead of trying to solve for the best scheduling and routing policies directly, we assume that we have access to a set of target values which represent the average age cost for every flow in the network. These target values could be application specific freshness requirements provided by a network administrator, or they could be the solution to an optimization program that optimizes some utility function of the average age costs. Given these targets, we set up a virtual queuing network that is stable if and only if there exists a feasible network control policy that can achieve these targets. Then, we use Lyapunov drift based methods to stabilize this system of virtual queues and achieve the desired target age costs.

In Section \ref{sec:model} we describe our system model. Then, we introduce notions of age debt and debt-stable scheduling policies in Section \ref{sec:age-debt}. We develop our heuristic scheduling and routing schemes using Lyapunov drift minimization techniques in Sections \ref{sec:drift} and \ref{sec:c_alpha}. Finally, we provide brief simulation results in Section \ref{sec:sim}.

\section{System Model}
\label{sec:model}
Consider a network with $N$ nodes connected by a fixed undirected graph $G(V,E)$. An edge $(i,j)$ means that nodes $i$ and $j$ can send packets to one another directly. We assume that at most one update can be sent over an edge in any given time-slot and takes exactly one time-slot to get delivered. We consider general interference constraints between the edges of the graph as well as unreliable transmissions. We further assume that all sources are active, i.e. they can generate fresh updates on demand and that there is no queuing at any node. Each node simply maintains a buffer for the freshest packet of each flow.

A flow is defined as a source node sending updates and a set of destination nodes that are interested in receiving updates from this source. Thus, each flow can be identified by its unique source node. Flows can be of three types depending on the the number of destination nodes: 
\begin{enumerate}
	\item \textbf{unicast}: the flow has a single destination node.
	\item \textbf{multicast}: the flow has multiple destination nodes, which are a strict subset of the remaining nodes.
	\item \textbf{broadcast}: every node other than the source itself is a destination node.
\end{enumerate}

We consider $K \leq N$ flows in this multi-hop network. For simplicity, we represent the source for the $k$th flow by $k$. The corresponding destination set for this flow is $D_k$. For every node $j$ that is a destination for $k$, we maintain an age of information process $A_{kj}(t)$ which tracks how old the information is at node $j$ about node $k$.
\begin{equation}
\label{eq:AoI_evolution_mh}
A_{kj}(t+1) =
\begin{cases}
\begin{aligned}
\min(A_{kj}(t),t - t_g) + 1, \text{if update generated}\\
\text{at time }t_g\text{ is delivered at time }t.
\end{aligned}\\
A_{kj}(t)+1, \text{if no new delivery at time }t.
\end{cases}
\end{equation}

We associate a monotone increasing age cost function for each source $k$ and corresponding destination $j \in D_k$ denoted by $f_{kj}(\cdot)$. Using these age cost functions, we maintain the effective age processes $B_{kj}(t) \triangleq f_{kj}(A_{kj}(t))$.

In general, a control policy needs to specify not only which links should be scheduled in each time-slot but also which flows should be transmitted along each link. We enumerate the set of all possible interference free choices of links and corresponding flows in the set $\mathcal{S}$. Thus, a member of set $\mathcal{S}$ contains a subset of links and corresponding flows which can be sent on these links in a single time-slot without interference. A valid network control policy must choose an action that is a member of the set $\mathcal{S}$ in every time-slot.

The typical goal of AoI-based scheduling and routing design for multihop networks is to minimize the time average of the expected age costs summed across flows:
\begin{equation}
\label{eq:age_cost_opt_mh}
\pi^{*} = \underset{\pi}{\operatorname{argmin}} \bigg( \lim_{T \rightarrow \infty}  \mathbb{E} \bigg[\frac{1}{T} \sum_{t = 1}^{T} \sum_{k=1}^{K} \sum_{j \in D_k} B_{kj}(t) \bigg] \bigg),
\end{equation}
where $\pi(t) \in \mathcal{S}, \forall t,\pi$.

Next, we introduce the notions of age-achievability and age debt virtual queues and show how stabilizing this network of virtual queues leads to minimization of AoI. 

\section{Age Debt}
\label{sec:age-debt}
We start by assuming that we have been given a target value of time average age cost for each source-destination pair; denoted by $\alpha_{kj}$ for the pair $(k,j)$. We aggregate the target values associated with each source-destination pair in the vector $\bm{\alpha}$. For any such target vector $\bm{\alpha}$, we define the notion of age-achievability below.
\begin{framed}
	\begin{definition}
		A vector $\bm{\alpha}$ is \textbf{age-achievable} if there exists a feasible network control policy $\pi$ such that		
		\begin{equation}
		\lim_{T \rightarrow \infty}  \frac{1}{T} \sum_{t = 1}^{T} B^{\pi}_{kj}(t)  \leq \alpha_{kj}, \forall j \in D_k, \forall k \text{ w.p. 1.}
		\end{equation}			
	\end{definition}
\end{framed}
In other words, a vector $\bm{\alpha}$ is age-achievable if the time-average of the effective age process for \textit{every} source-destination pair $(k,j)$ is upper bounded by the target value $\alpha_{kj}$, under some feasible network control policy. 

Note that the combination of general cost functions and achievability targets allows us to capture very general freshness requirements which might be useful in practical system specifications. For example, if an application requires that the empirical distribution of the age process $A_{kj}(t)$ should satisfy $\mathbb{P}(A_{kj}(t) \geq M) \leq \epsilon$, then we can capture this by setting the cost function $f_{kj}(h) = \bm{1}_{h \geq M}$ and the corresponding target to be $\alpha_{kj} = \epsilon$. 

We now define a set of virtual queues called age-debt queues for every source-destination pair $(k,j)$. These queues measure how much the effective age process exceeds its target value $\alpha_{kj}$, summed over time. Our definition of debt is inspired by the notion of throughput debt as introduced in \cite{hou2009theory}.
\begin{framed}
	\begin{definition}
		Given a target vector $\bm{\alpha}$, the \textbf{age debt queue} for source-destination pair $kj$ at time $t$ under a policy $\pi$, given by $Q^{\pi}_{kj}(t)$, evolves as 		
		\begin{equation}
        \begin{aligned}
        Q^{\pi}_{kj}(t+1) = \bigg[Q^{\pi}_{kj}(t) + B^{\pi}_{kj}(t+1) - \alpha_{kj}\bigg]^{+}, \forall j \in D_k, \\ \text{ and } \forall k \in \{1,...,K\}.
        \end{aligned}
        \end{equation}
		To complete the definition, each age debt queue starts at zero, i.e. $Q^{\pi}_{kj}(0) = 0, \forall j,k$.		
	\end{definition}
\end{framed}
We now introduce a notion of stability for these age debt queues. This is similar to how rate stability is typically defined in queueing networks \cite{neely2010stability}.
\begin{framed}
	\begin{definition}
		We say that the network of age debt queues is \textbf{stable} under a policy $\pi$ and a given target vector $\bm{\alpha}$ if the following condition holds:
		\begin{equation}
		\lim_{T \rightarrow \infty} \mathbb{E} \bigg[\sum_{k=1}^{K} \sum_{j \in D_k} \frac{Q^{\pi}_{kj}(T)}{T} \bigg] = 0,
		\end{equation}
		where the expectation is taken over the randomness in the channel processes.
	\end{definition}
\end{framed}
We also establish an equivalence relationship between age-achievability of a vector $\bm{\alpha}$ and the stability of the corresponding network of age debt queues.
\begin{lemma}
		\label{lem:debt_equi}
		A target vector $\bm{\alpha}$ is age-achievable \textit{if and only if} there exists a network control policy $\pi$, possibly dependent on $\bm{\alpha}$, that stabilizes the network of source-destination age debt queues.
\end{lemma}
\begin{IEEEproof}
   See Appendix \ref{pf:lem_equi}.
\end{IEEEproof}
	

Next, we define a debt-stable scheduling policy. Such a policy takes a target vector $\bm{\alpha}$ as an input and stabilizes the network of corresponding age debt queues.
\begin{framed}
	\begin{definition}
		A \textbf{debt-stable} scheduling policy $\pi$ stabilizes the set of age-debt queues for any given target vector $\bm{\alpha}$ that is age-achievable. 
	\end{definition}
\end{framed}

The notions introduced until now effectively allow us to convert the minimum age cost problem introduced in \eqref{eq:age_cost_opt_mh} into a network stability problem. Suppose $\pi^{*}$ is a solution to the optimization problem \eqref{eq:age_cost_opt_mh}. Further, suppose that the time average of the $kj$th effective age process under $\pi^{*}$ is given by 
\begin{equation}
	\lim_{T \rightarrow \infty}  \mathbb{E} \bigg[\frac{1}{T} \sum_{t = 1}^{T} B^{\pi^{*}}_{kj}(t) \bigg] = \alpha^{*}_{kj}, \forall (k,j).
\end{equation}
Clearly, if we have oracle access to an optimal age cost vector $\bm{\alpha^{*}}$ and know how to design a debt-stable policy then we can perform minimum age cost scheduling. If the debt-stable policy is much lower in computational complexity than solving \eqref{eq:age_cost_opt_mh} directly, then we can also solve \eqref{eq:age_cost_opt_mh} at the same lower complexity (assuming oracle access to $\bm{\alpha^{*}}$). We now discuss a heuristic approach to designing debt-stable policies.

\section{Lyapunov Drift Approach}
\label{sec:drift}
\subsection{Single-Hop Broadcast}
\label{sec:single_hop}
We first consider the special case of single-hop broadcast networks. This setting is easier to analyze since it only requires scheduling and no routing and it also highlights key structural properties of our proposed policy.

Consider a $N$ node star network where each of the nodes $1,...,N-1$ has an edge to node $N$. These nodes wish to send packets to the central node $N$. Due to broadcast interference constraints, only one node can transmit in any given time-slot. Since the destination for every flow is $N$, we can drop the destination in our notation. The age evolution is given by 
\begin{equation}
\label{eq:AoI_evolution}
A^{\pi}_i(t+1) =
 \begin{cases}
      A^{\pi}_i(t)+1, & \text{if } i \notin \pi(t) \text{ or } c_{i}(t) = 0 \\
      1, &  \text{if }i \in \pi(t) \text{ and } c_i(t) = 1. 
 \end{cases}
\end{equation}
Here $\pi(t)$ is the source scheduled in time-slot $t$ and $c_i(t)$ is an indicator variable denoting edge reliability between node $i$ and node $N$ at time $t$. Given an age-cost function $f_i(A_i(t))$ and a corresponding target value $\alpha_i$, the debt queue evolution for node $i$ is given by:
\begin{equation}
\label{eq:s_debt}
Q^{\pi}_i(t+1) = \bigg[Q^{\pi}_i(t) + f_i(A^{\pi}_i(t+1)) - \alpha_i\bigg]^{+}. 
\end{equation}

Given a target vector $\bm{\alpha}$, we will use a Lyapunov drift based scheduling scheme to try and achieve debt stability. To do so, we first define a Lyapunov function for our system of virtual queues:
\begin{equation}
	L(t) \triangleq \sum_{i=1}^{N-1} Q^{2}_i(t).
\end{equation} 

Using this Lyapunov function, we then define the \textit{age debt scheduling policy} $\pi^{\text{AD}}$ as:
\begin{equation}
\label{eq:age_debt_policy}
\pi^{\text{AD}}(t) = \underset{a \in \mathcal{S}}{\operatorname{argmin}} \bigg( \mathbb{E} \big[ L(t+1) - L(t)\big] \bigg),
\end{equation}
where the expectation is taken over the randomness in channel reliabilities $\bm{c}(t)$. The following lemma describes what the drift minimizing actions look like in practice.
\begin{lemma}
\label{lem:drift_exp}
Suppose that the links between each source $i$ and the destination $N$ are i.i.d. Bernoulli w.p. $p_i$ in every time-slot. Further, if each age cost function $f_i(\cdot)$ is upper bounded by a large constant $D$, then the policy $\pi(t)$ below minimizes an upper bound on the Lyapunov drift in every time-slot.
\begin{equation}
    \pi(t) = \underset{i \in 1,...,N-1}{\operatorname{argmax}} \bigg( p_i Q_i(t) \big(f_i(A_i(t)+1) - f_i(1)\big) \bigg).
\end{equation}
\end{lemma}
\begin{IEEEproof}
   See Appendix \ref{pf:drift}.
\end{IEEEproof}
In other words, a drift minimizing policy chooses the source with the largest product of link reliability, current age debt and current age cost. This structure of the drift minimizing policy can be contrasted with the max-weight policy proposed in \cite{kadota2018scheduling} which chooses the source with the largest value of $p_i w_i A_i(t) (A_i(t)+2)$ given weights $w_i$. Similarly, the Whittle index policy proposed in \cite{tripathi2019whittle}, chooses the source with the largest value of $W_i(A_i(t))$, where $W_i(\cdot)$ is Whittle-index corresponding to the age cost $f_i(\cdot)$.

Note that to compute $\pi^{\text{AD}}(t)$, the scheduler needs to iterate over the set of sources only once. So the per slot computational complexity of this policy grows linearly in $N$. This is similar to the complexity of the Whittle index policy proposed in \cite{kadota2018scheduling,tripathi2019whittle} and the max-weight policies proposed in \cite{kadota2018scheduling,talak2018optimizing}. By contrast, a dynamic programming approach to solve \eqref{eq:age_cost_opt_mh} directly has per slot computational complexity that grows exponentially in $N$. This highlights the key strength of our approach. If the scheduler has some way to set the targets for each source optimally, then the age debt policy is a good low complexity heuristic for age minimization.

\subsection{General Networks}
\label{sec:multihop}
The general multihop setting is more challenging. Simply using one-slot Lyapunov drift to try and achieve debt stability does not work directly in the multihop setting. We highlight this with a simple example.

\begin{figure}
	\centering
	\includegraphics[width=0.7\linewidth]{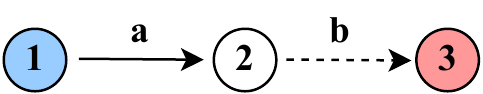}
	\caption{}
	\label{fig:two_hop}
\end{figure}

Consider the three node network described in Figure~\ref{fig:two_hop} with a single unicast flow from node $1$ to node $3$. The interference constraint enforces that only one of the two edges $a$ and $b$ can be activated in any time-slot. Suppose that we are interested in minimizing the time average of the age process $A_{13}(t)$. Given a target value $\alpha_{13}$, we set up the age debt queue as follows:
\begin{equation}
	Q^{\pi}_{13}(t+1) = \bigg[Q^{\pi}_{13}(t) + A^{\pi}_{13}(t+1) - \alpha_{13}\bigg]^{+}.
\end{equation}

We will try to use the one slot Lyapunov drift minimizing policy to stabilize $Q_{13}(t)$ in this network. To do so, we solve the following optimization in every time-slot:
\begin{equation}
\pi^{\text{AD}}(t) = \underset{x \in \{a,b\} }{\operatorname{argmin}} \bigg( \mathbb{E} \big[ Q_{13}^{2}(t+1) - Q_{13}^{2}(t)\big] \bigg).
\end{equation}

At $t=1$, activating either edge $a$ or edge $b$ has no effect on the debt $Q_{13}(2)$ since node $2$ does not have any packet from node $1$. If we break ties in favour of edge $b$, then it is activated but no new packet is delivered to node $3$. At $t=2$, since node $2$ still does not have any new update from node $1$, no action taken can affect the debt $Q_{13}(3)$. Using the same tie-break rule, we would again schedule edge $b$. This process keeps on repeating and the age debt queue $Q_{13}$ blows up irrespective of the value of $\alpha_{13}$, even though the age optimal policy in this setting is to simply alternate between $a$ and $b$ in every time-slot.

The example above illustrates why one-slot Lyapunov drift based techniques fail in stabilizing debt queues in multihop networks. The policy designer using Lyapunov drift is constrained to optimizing \textit{only one time-step into the future}. So, if every possible scheduling and routing action has no effect on the age debt queues in the immediate next time-slot, the one step drift minimizing procedure and does not provide any information on which action should be chosen to stabilize the network.   

This suggests that to be able to use one-slot drift minimizing techniques for stability there should be a virtual queue for every intermediate node that tracks both the current age debt at the destination and the potential reduction in debt at the destination upon forwarding a fresh packet. If we can set up such queues, then large values of debt at intermediate nodes would lead to fresh packets being sent to the next hops via one-slot drift minimizing actions, eventually reaching the destination and stabilizing the age debt queues. 

Let $Q_{kj}^{i}(t)$ denote such a debt queue corresponding to flow $(k,j)$ at an intermediate node $i$. These additional queues at every intermediate node combined with the original debt queues form our virtual network. The Lyapunov function that we use for scheduling and routing is given by:
\begin{equation}
	L(t) \triangleq \sum_{k=1}^{K} \sum_{j \in D_k} \bigg(Q^{2}_{kj}(t) + \sum_{i \notin D_k, i \neq k} (Q^{i}_{kj}(t))^2 \bigg)
\end{equation}

The Age Debt scheduling and routing policy is to choose the activation set and corresponding flows that minimizes the expected Lyapunov drift.
\begin{equation}
\pi^{\text{AD}}(t) = \underset{a \in \mathcal{S}}{\operatorname{argmin}} \bigg( \mathbb{E} \big[ L(t+1) - L(t)\big] \bigg),
\end{equation}
where the expectation is taken over the randomness in channel reliabilities $\bm{c}(t)$.

\subsection{Intermediate Debt Queues}
We now discuss how to set up the age debt queues $Q_{kj}^{i}(t)$ for intermediate nodes to augment the original network of queues. Note that there are no intermediate nodes for broadcast flows since every node other than the source is a destination. 

Consider a source-destination pair $(k,j)$ for a unicast/multicast flow $k$ and an intermediate node $i$ that is not a destination for the flow originating at $k$. We want to set up the age debt queue $Q_{kj}^{i}(t)$ at $i$ for the pair $(k,j)$. We maintain an age process for flow $k$ at node $i$, even though there is no associated cost or target value for this age process.  
\begin{equation}
A_{ki}(t+1) =
\begin{cases}
\begin{aligned}
\min(A_{ki}(t),t - t_g) + 1, \text{if update generated}\\
\text{at time }t_g\text{ is delivered at time }t.
\end{aligned}\\
A_{ki}(t)+1, \text{if no new delivery at time }t.
\end{cases}
\end{equation}
Here $A_{ki}(t)$ measures how old the information at node $i$ is regarding node $k$. We split the debt queue's evolution into two cases. 

\textbf{Case 1}: When node $i$ forwards a flow $k$ packet on a set of adjacent links $L$. Let $h^{L}_{ij}$ be the minimum number of hops it takes to reach node $j$ from node $i$, where the first hop can only include edges in the set $L$. Here, $h^{L}_{ij}$ measures the minimum delay with which the packet that was forwarded by $i$ gets delivered at $j$. The age debt queue $Q_{kj}^{i}(t)$, when node $i$ is forwarding a flow $k$ packet along the link set $L$, evolves as:
\begin{equation}
\label{eq:c1}
\begin{aligned}
Q^{i}_{kj}(t+1) = \bigg[Q^{i}_{kj}(t) + f_{kj}\big(\min \{A_{ki}(t),A_{kj}(t) \} \\ + h^{L}_{ij}\big) - \alpha_{kj}\bigg]^{+}.
\end{aligned}
\end{equation}
This measures the most optimistic change in age debt possible at the destination using the current packet transmission from node $i$.

\textbf{Case 2:} When node $i$ does not forward a packet from node $k$ along any of its adjacent edges, then the age debt queue evolves as below.
\begin{equation}
\label{eq:c2}
Q^{i}_{kj}(t+1) = \bigg[Q^{i}_{kj}(t) + B_{kj}(t+1) - \alpha_{kj}\bigg]^{+}.
\end{equation}
This means that the intermediate queue simply tracks the change in debt at the destination when it is not forwarding a relevant packet. If the destination is not receiving fresh packets from anywhere in the network then this would increase the intermediate debt queue.

Thus, the debt at an intermediate node $i$ for a source-destination pair $(k,j)$ blows up if (a) either the destination has not received fresh packets for a long time and node $i$ did not forward any packets from $k$ (i.e. \eqref{eq:c2}) or if (b) node $i$ keeps forwarding stale packets from $k$ (i.e. \eqref{eq:c1}). A drift minimizing policy will then try to ensure that either the destination debt queue is small, or node $i$ forwards fresh packets of flow $k$ towards the destination.

\section{Choosing Target Vectors}
\label{sec:c_alpha}
In the preceding sections, we have developed a general framework of age achievability where given a target average age cost for every source-destination pair, we formulate a corresponding network stability problem and attempt to solve it via one slot Lyapunov drift minimization. In this section, we discuss how to choose the right target vectors, such that they lead to minimum sum age cost.

In the absence of an optimization oracle that provides access to $\bm{\alpha^{*}}$ or a system administrator who specifies average age cost targets based on the underlying application requirements, we develop a simple heuristic to dynamically update $\bm{\alpha}$ in order to optimize utility based on the state of the underlying debt queues.



The following optimization problem needs to be solved to find the best target vector $\bm{\alpha^{*}}$. 
\begin{equation}
\label{eq:alpha_opt}
\begin{aligned}
\underset{\bm{\alpha}}{\operatorname{argmin}} & ~\bigg( \sum_{k=1}^{K} \sum_{j \in D_k} \alpha_{kj} \bigg),\\
\text{s.t. } & \bm{\alpha} \text{ is age-achievable}. 
\end{aligned}
\end{equation}
Note that this problem has the same optimal value as \eqref{eq:age_cost_opt_mh}.

\subsection{Gradient Descent}
We want to use a gradient descent like approach to solve \eqref{eq:alpha_opt} and find $\bm{\alpha^{*}}$. The problem with doing so is that we do not have a simple characterization of the age-achievability region or a low complexity method to test whether a vector is achievable or not. In fact, we do not even know if the region is convex.

To resolve this, we use Lemma \ref{lem:debt_equi}. If the network of source-destination age debt queues is unstable for a given value of $\bm{\alpha}$, then $\bm{\alpha}$ lies outside the age-achievability region. This immediately suggests the following gradient descent like algorithm. 

\begin{algorithm}
	\DontPrintSemicolon
	\SetKwInOut{Input}{Input}\SetKwInOut{Output}{Output}
	\Input{epoch size $W$, number of epochs $E$, step-size $\eta > 0$, threshold $\epsilon > 0$, initialization $\bm{\alpha}^{(1)}$}
	\BlankLine
	\While{ $e \in 1,...,E$ }{
		Set up age debt queues using $\bm{\alpha}^{(e)}$ and initialize each queue to 0\;
		\While{ $t \in 1,..., W$}{
		Schedule and route using age debt $\pi^{\text{AD}}(t) = \underset{a \in \mathcal{S}}{\operatorname{argmin}} \bigg( \mathbb{E} \big[ L(t+1) - L(t)\big] \bigg)$,
		}		
		\If{$\exists (k,j)$ s.t. $Q_{kj}(W) > \epsilon W$}{
		Increase target values for unstable queues:\\	
		$\alpha^{(e+1)}_{kj}  = \alpha^{(e)}_{kj} + \eta$, $\forall (k,j)$ s.t. $Q_{kj}(W) > \epsilon W$\;
		Other targets remain unchanged:\\
		$\alpha^{(e+1)}_{kj}  = \alpha^{(e)}_{kj}$, $\forall (k,j)$ s.t. $Q_{kj}(W) \leq \epsilon W$\;
		}		
		\Else{
			Update all target values using gradients:	
			$\alpha^{(e+1)}_{kj}  = \alpha^{(e)}_{kj} - \eta$, $\forall (k,j)$.\;
		}
	}
	\caption{Age Debt - Gradient Descent}
	\label{alg:ADGD}
\end{algorithm}

The algorithm above runs the age debt policy for epochs of length $W$ time-slots. Within an epoch the target vector remains fixed. At the end of the epoch, we use the value of the source-destination age debt queues to update the corresponding targets. If the network has at least one queue with debt larger than a threshold, it suggests that the current vector is not achievable. So, we increase the values of $\bm{\alpha}$ for the sources with large values of debt. If the network has all queues with debt below a threshold, the current vector is likely achievable. So, we update the entire target vector using gradient descent. Note that this approach takes a large number of time-slots to converge to a good candidate target vector $\bm{\alpha}$.

\subsection{Flow Control}
Another way to dynamically set the target vectors is to take a flow control approach for solving the optimization problem \eqref{eq:alpha_opt}, similar to \cite{georgiadis2006resource}. Algorithm \ref{alg:ADFC} describes the details.
\begin{algorithm}
	\DontPrintSemicolon
	\SetKwInOut{Input}{Input}\SetKwInOut{Output}{Output}
	\Input{parameter $V>0$, upper bound $\alpha_{\text{max}}$, initialization $\bm{\alpha}^{(1)}$}
	\BlankLine
	\While{ $t \in 1,...,T$ }{
		Use $\bm{\alpha}^{(t)}$ to update debt queue values at time $t$ \;
		Update $\bm{\alpha}$ by solving the optimization below:\\
		\begin{equation*}
		\bm{\alpha}^{(t+1)} = 
			\begin{aligned}
			\underset{\bm{\alpha}}{\operatorname{argmin}} & \bigg( \sum_{k=1}^{K} \sum_{j \in D_k} V \alpha_{kj} - \alpha_{kj} Q_{kj}(t) \bigg)  ,\\
			\text{s.t.  } & \bm{\alpha} \geq 1, \bm{\alpha} \leq \alpha_{\text{max}}. 
			\end{aligned}
		\end{equation*}\;
		Use $\bm{\alpha}^{(t+1)}$ to compute the scheduling and routing decision that minimizes drift:
		$\pi(t) = \underset{a \in \mathcal{S}}{\operatorname{argmin}} \bigg( \mathbb{E} \big[ L(t+1) - L(t)\big] \bigg)$\;						
	}		
	\caption{Age Debt - Flow Control}
	\label{alg:ADFC}
\end{algorithm}

The flow control based age debt policy tries to tradeoff between the stability of the queueing network and the optimization of targets using a parameter $V > 0$. In every time-slot, the flow control optimization sets the target $\bm{\alpha}$ for the next time-slot and then the scheduling and routing decisions are computed by minimizing Lyapunov drift.

The update optimization in step 4 of Algorithm \ref{alg:ADFC} can be simplified to the rule below:
\begin{equation}
	\alpha^{(t+1)}_{kj} = \begin{cases}
	\alpha_{\text{max}}, &\text{ if } Q_{kj}(t) > V\\
	1, &\text{ if }Q_{kj}(t) \leq V,
	\end{cases}
	\forall (k,j).
\end{equation}
Thus, instead of converging to a target vector as in the case with gradient descent, 
the flow control approach dynamically switches the value of targets in every time-slot. This means we do not need to wait a long period of time for convergence. When current debts are high, future targets are set to be high pushing the debts lower. Similarly, when the current debts are low, future targets are also set low, pushing the debts higher. The parameter $V$ decides the threshold between high and low values of the debt queues.


\section{Simulations}
\label{sec:sim}
\begin{figure}
	\centering
	\includegraphics[width=0.95\linewidth]{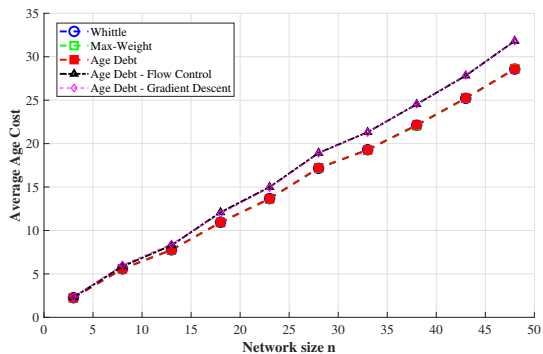}
	\caption{Weighted-sum AoI minimization in broadcast networks with unreliable channels}
	\label{fig:single_lin_uch}
\end{figure}
First, we consider the weighted-sum AoI problem in broadcast networks with unreliable channels. There are $N$ nodes in the network and the weight of the $i$th node $w_i$ is set to $i/N$. Link connection probabilities are chosen uniformly from the set $[0.6,1]$. Figure \ref{fig:single_lin_uch} plots the performance of the age debt policy along with the max-weight and Whittle index policies proposed in \cite{kadota2018scheduling} which are known to be close to optimal. We observe that when the age debt policy is provided the max-weight average cost as the target vector, it replicates near optimal performance. Further, the flow control and gradient descent versions of age debt only have a small gap to the max-weight/Whittle policies despite not having access to $\bm{\alpha}$ beforehand. 

\begin{figure}
	\centering
	\includegraphics[width=0.95\linewidth]{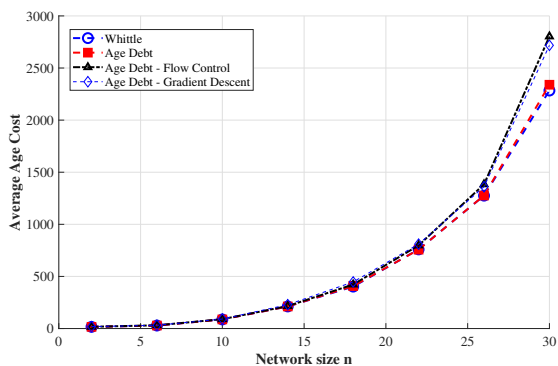}
	\caption{Functions of Age minimization in broadcast networks with reliable channels}
	\label{fig:single_func_rch}
\end{figure}
Next, we consider general functions of age minimization in the single-hop wireless broadcast setting. There are $N$ nodes in the network and the cost of AoI for each node is chosen from the set of functions $\{15A(t), e^{A(t)}, (A(t))^2 \text{ and }  (A(t))^3\}$. Figure \ref{fig:single_func_rch} plots the performance of the age-debt policy and its variants along with the Whittle index policy proposed in \cite{tripathi2019whittle}. As for the linear AoI case, we observe that age debt is able to replicate the Whittle policy's performance when provided its average cost as the target vector. The flow control and gradient descent variants are also only a small gap away in performance without knowing $\bm{\alpha}$ beforehand.

We also look at the functions of age problem with $N=4$ in more detail. The age cost functions for each node are as follows $f_1(A_1(t)) = 15A_1(t)$, $f_2(A_2(t)) = e^{A_2(t)}$, $f_3(A_3(t)) = (A_3(t))^2$ and $f_4(A_4(t)) = (A_4(t))^3$. First, we use dynamic programming to compute the optimal policy $\pi^{*}$ which minimizes average age cost. The time average age costs under this policy are given by $\alpha^{*}_1 = 45.0, \alpha^{*}_2 = 14.52, \alpha^{*}_3 = 17.20,$ and $\alpha^{*}_4 = 11.0$,  while the total sum cost is 87.72. Using these as target values, we set up debt queues and implement the age-debt policy. 

\begin{figure}
	\centering
	\includegraphics[width=0.95\linewidth]{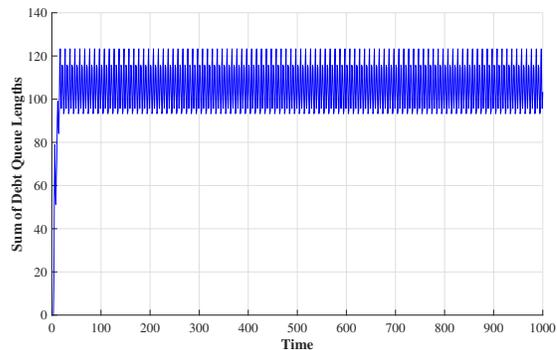}
	\caption{Sum of virtual debt queues vs time}
	\label{fig:dql}
\end{figure}
Figure \ref{fig:dql} plots the sum of the 4 age debt queues $\sum_{i=1}^{4}Q_i(t)$ under the age-debt policy implemented using the optimal $\bm{\alpha}^{*}$ from above. We observe that the age debt policy indeed stabilizes the debt queues since queue lengths don't grow with time. As a corollary, it also achieves age cost optimality in this setting. On the other hand, the Whittle index policy from \cite{tripathi2019whittle} achieves a total sum cost of 88.34, a fixed but small distance away from the optimal cost of 87.72. This suggests that age-debt might be a way to achieve exact optimality instead of near optimality when access to $\bm{\alpha}^{*}$ is available.

\begin{figure}
	\centering
	\includegraphics[width=0.95\linewidth]{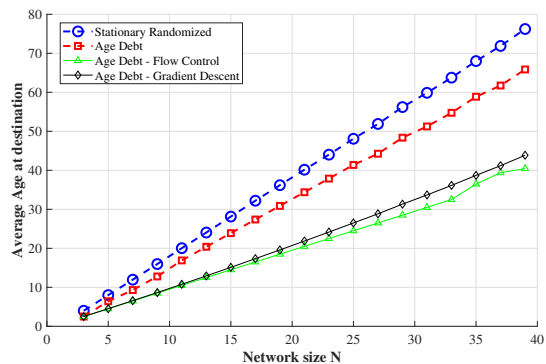}
	\caption{Age minimization of a single unicast flow in line networks (neighboring nodes interfere)}
	\label{fig:unicast_line}
\end{figure}
Next, we consider scheduling for a single unicast flow on the line network, as studied in \cite{talak2017minimizing}. Consider $N$ nodes arranged in a line network from 1 to $N$. Node $1$ wants to sent packets to node $N$, however not all nodes can transmit simultaneously. We consider a simple interference constraint - in any given time-slot either all even numbered nodes or all odd numbered nodes can forward packets. This ensures that no two adjacent nodes send interfering transmissions. Figure \ref{fig:unicast_line} plots the performance of age-debt and its flow-control and gradient-descent variations along with the optimal stationary randomized policy proposed in \cite{talak2017minimizing}. We observe that age-debt outperforms the stationary randomized policy despite using its average costs as the target vector. The dynamic variants of age-debt significantly outperform the stationary randomized policy. We also note that the gap in performance would increase in settings where multiple paths are available which age-debt can utilize for routing, unlike the stationary randomized approach.

\begin{figure}
	\centering
	\includegraphics[width=0.95\linewidth]{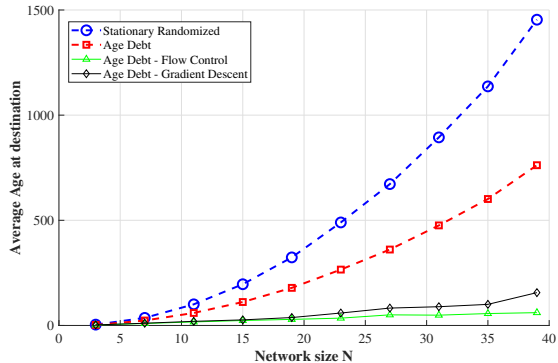}
	\caption{Age minimization of a single unicast flow in line networks (all nodes interfere)}
	\label{fig:unicast_line_s}
\end{figure}
We also consider a different kind of interference constraint in the same line network example. Now, all nodes interfere with one another, and only one node can transmit successfully in any given time-slot. We plot the performance of the optimal stationary randomized policy along with age-debt and its variants against the number of nodes in the system in Figure \ref{fig:unicast_line_s}. We again observe a large gap in performance between the optimal randomized policy and our proposed methods. This is consistent with the performance bounds in \cite{talak2017minimizing}, where it was proved that the best stationary randomized policy can have performance that is a constant factor away from optimal but the factor grows exponentially in the size of the network (in the worst case).

\begin{figure}
	\centering
	\includegraphics[width=\linewidth]{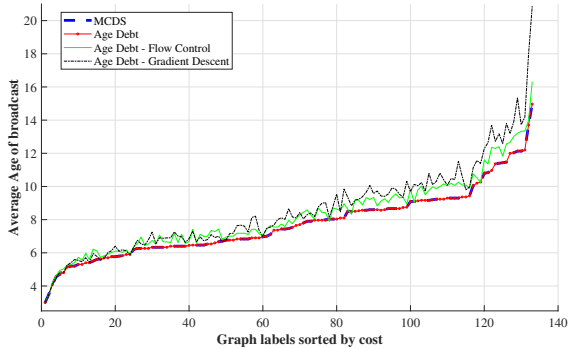}
	\caption{Age minimization of broadcast flows in multihop networks with 5 and 6 nodes}
	\label{fig:multi_equal}
\end{figure}
Finally, we consider average age minimization for all-to-all broadcast flows in multihop networks similar to \cite{farazi2018age}. We consider all possible connected network topologies with 5 or 6 nodes (a total of 133 graphs). Figure \ref{fig:multi_equal} plots the performance of the age-debt policy and its variants along with the near optimal minimum connected dominating set (MCDS) based scheme proposed in \cite{farazi2018age} for each of these networks. The x-axis represents the graph labels numbered from 1 to 133, sorted according to the average age achieved by the MCDS scheme.

We observe that age-debt achieves the same performance as the MCDS scheme when provided its average cost as the target vector. Further, age-debt with flow control achieves performance that is very close to that of the MCDS scheme without requiring knowledge of $\bm{\alpha}$.

\begin{figure}
	\centering
	\includegraphics[width=\linewidth]{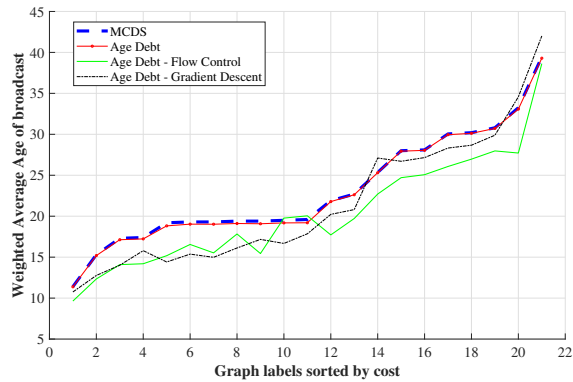}
	\caption{Weighted Age minimization of broadcast flows in multihop networks with 5 nodes}
	\label{fig:multi_w}
\end{figure}
We also consider the same broadcast setting but now with weighted-sum AoI as the minimization objective instead of just AoI. We consider all possible connected graphs with 5 nodes (21 in total). We set the importance weight of one node to $15$ (giving it a higher priority) and the rest of the 4 nodes to $1$. Figure \ref{fig:multi_w} plots the performance of the MCDS scheme along with age-debt and its variants. As expected, age-debt policy replicates the performance of the MCDS scheme since it is provided the average age-cost realized by MCDS as the target. Interestingly, flow-control outperforms MCDS since it is able to adapt to a better target $\bm{\alpha}$ in the presence of weights and asymmetry. This is consistent with the fact that the MCDS scheme is not designed for minimizing weighted-sum AoI. It also highlights the relative ease with which age-debt can be adapted to weights and general AoI cost functions.

Note that the complexity of implementing the flow-control scheme is polynomial in the network size per time-slot. This suggests that age-debt and its variants are a good candidate for low complexity near optimal age scheduling in general networks.

We also observe that the flow control variant of age-debt is the method of choice in the absence of known $\bm{\alpha}$. The gradient descent variant has parameters that are hard to configure for networks of different sizes and takes a long time to converge. The flow-control method has just two parameters $V$ and $\alpha_{\text{max}}$ and does not require time for convergence.

Interesting directions of future work involve proving performance bounds on the age debt policy and its variants, implementing age debt in a distributed fashion, and considering stochastic arrivals and time-varying topologies in the underlying network.

\bibliographystyle{ieeetr}
\bibliography{bibliography_2}

\appendix
\subsection{Proof of Lemma \ref{lem:debt_equi}}
\label{pf:lem_equi}
We will prove this under the assumption that the AoI cost functions $f_{kj}(\cdot)$ are upper-bounded by a fixed constant $D$ for every source-destination pair $(k,j)$. This is a mild assumption because $D$ can be set to a very high value (in the order of years) which will never be attained in practical systems under any reasonable policy.

We note that the arrival process to the debt queue $Q_{kj}(t)$ is given by the effective age process $B_{kj}(t)$, while the departures in every time-slot are just $\alpha_{kj}$. Using the boundedness assumption, both arrivals and departures are strictly upper-bounded by $D$. The result immediately follows from Theorem 2(c) in \cite{neely2010stability} which relates mean-rate stability of a queue to time-averages of the arrival and departure processes. 

\subsection{Proof of Lemma \ref{lem:drift_exp}}
\label{pf:drift}
The debt queues in this setting evolve as follows:
\begin{equation}
    Q_i(t+1) = \bigg[Q_i(t) + f_i(A_i(t+1)) - \alpha_i\bigg]^{+}, \forall i.
\end{equation}
The AoI evolves as:
\begin{equation}
A_i(t+1)
\begin{cases}
      A_i(t)+1, & \text{if } i \notin \pi(t) \text{ or } c_{i}(t) = 0 \\
      1, &  \text{if }i \in \pi(t) \text{ and } c_i(t) = 1. 
 \end{cases}
\end{equation}
Here $c_i(t) = 1$ i.id. with probability $p_i$ in every time-slot.

Let $\Delta(t) \triangleq L(t+1) - L(t)$. Then,
\begin{equation}
\begin{split}
    \mathbb{E}[\Delta(t)] & = \sum_{i} \mathbb{E}\bigg[ (Q_i(t+1))^2 - (Q_i(t))^2 \bigg]\\
    & \leq \sum_i \mathbb{E}\bigg[ \alpha_i^2 - 2\alpha_i Q_i(t) + (f_i(A_i(t+1)))^2 + \\ & 2Q_i(t)f_i(A_i(t+1)) - 2\alpha_i f(A_i(t+1)) \bigg]\\
    & \leq \sum_i \bigg[ D^2 + 2 Q_i(t) (\mathbb{E}[f_i(A_i(t+1))] - \alpha_i) \bigg]
\end{split}
\end{equation}

The first inequality follows from the evolution of debt queues. The second inequality follows from the boundedness assumption on $f_{i}(\cdot)$, i.e. $f_i(h) \leq D, \forall h$. Now, we will minimize the RHS of the expression above. We can drop the term $D^2$ since it is a constant.
\begin{equation}
\begin{split}
    &\underset{\pi(t) \in 1,...,N-1}{\operatorname{argmin}} \sum_i  Q_i(t) \big(\mathbb{E}[f_i(A_i(t+1))] - \alpha_i\big) \\
    = & \underset{\pi(t) \in 1,...,N-1}{\operatorname{argmin}} \sum_i  Q_i(t) \mathbb{E}[f_i(A_i(t+1))] \\
    = & \underset{j \in 1,...,N-1}{\operatorname{argmin}} \bigg[\sum_i \bigg( Q_i(t) f_i(A_i(t)+1) \bigg) + \\
    & ~~~~~~p_j Q_j(t) (f_j(1) - f_j(A_j(t) + 1))\bigg]\\
    = & \underset{j \in 1,...,N-1}{\operatorname{argmax}} \bigg[ p_j Q_j(t) \big(f_j(A_j(t) + 1) - f_j(1)\big) \bigg]
\end{split}
\end{equation}
The first equality follows since $Q_i(t)\alpha_i$ does not depend on the scheduling decision $\pi(t)$. The second equality follows from the evolution of AoI given $\pi(t)=j$. The third equality follows since the summation term does not depend on the scheduling choice $j$. This completes the proof.
\end{document}